\documentclass{article}

\usepackage{arxiv}

\usepackage{subcaption}
\usepackage{adjustbox}
\newtheorem{remark}{Remark}
\usepackage[noadjust]{cite}
\usepackage{graphicx}
\usepackage{amssymb}
\usepackage{epstopdf}
\usepackage[english]{babel}
\usepackage{amsmath}
\usepackage{amsfonts}
\usepackage[usenames,dvipsnames,svgnames,table]{xcolor} 
\def\qed{\hfill \mbox{\rule[0pt]{1.5ex}{1.5ex}}}
\def\R{\mathbb{R}}

\def\cM{{\cal M}}

\newtheorem{prop}{Proposition}

\newcommand{\be}{\begin{equation}}
\newcommand{\ee}{\end{equation}}
\newcommand{\bc}{\begin{center}}
\newcommand{\ec}{\end{center}}

\title{Embedding classic chaotic maps in simple discrete-time memristor circuits}

\newif\ifuniqueAffiliation

\ifuniqueAffiliation 
\author{ \href{https://orcid.org/0000-0000-0000-0000}{\includegraphics[scale=0.06]{orcid.pdf}\hspace{1mm}David S.~Hippocampus}\thanks{Use footnote for providing further
		information about author (webpage, alternative
		address)---\emph{not} for acknowledging funding agencies.} \\
	Department of Computer Science\\
	Cranberry-Lemon University\\
	Pittsburgh, PA 15213 \\
	\texttt{hippo@cs.cranberry-lemon.edu} \\
	\And
	\href{https://orcid.org/0000-0000-0000-0000}{\includegraphics[scale=0.06]{orcid.pdf}\hspace{1mm}Elias D.~Striatum} \\
	Department of Electrical Engineering\\
	Mount-Sheikh University\\
	Santa Narimana, Levand \\
	\texttt{stariate@ee.mount-sheikh.edu} \\
}
\else
\usepackage{authblk}

\author[1]{{Mauro Di Marco\thanks{\texttt{mauro.dimarco@unisi.it}}}}
\author[1]{{Mauro Forti\thanks{\texttt{forti@dii.unisi.it}}}}
\author[1]{{Luca Pancioni \thanks{\texttt{pancioni@dii.unisi.it}}}}
\author[2]{Giacomo Innocenti\thanks{\texttt{giacomo.innocenti@unifi.it}}}
\author[2]{Alberto Tesi\thanks{\texttt{alberto.tesi@unifi.it}}}
\affil[1]{Department of Information Engineering and Mathematics,
University of Siena, via Roma 56 - 53100 Siena, Italy.}
\affil[2]{Department of Information Engineering,
University of Florence, via S. Marta 3 - 50139 Firenze, Italy.}
\fi

\renewcommand{\headeright}{}
\renewcommand{\undertitle}{}
\renewcommand{\shorttitle}{Embedding chaotic maps in discrete-time memristor circuits}

\begin{document}

\maketitle

\begin{abstract}
In the last few years the literature has witnessed a remarkable surge of interest for
maps implemented by discrete-time (DT) memristor circuits. One main reason is that from
numerical simulations it appears that
even for simple memristor circuits the maps can easily display complex dynamics, including chaos and hyperchaos, which are of relevant interest for engineering applications. Goal of this manuscript is to investigate on the reasons underlying this type of complex behavior. To this end, the
manuscript considers the map implemented by the simplest memristor circuit given by a
capacitor and an ideal flux-controlled memristor or an inductor and an ideal charge-controlled
memristor. In particular, the manuscript uses the DT flux-charge analysis method (FCAM) introduced
in a recent paper to ensure that the first integrals and foliation in invariant manifolds
of continuous-time (CT) memristor circuits are preserved exactly in the discretization for any step size.
DT-FCAM yields a two-dimensional map in the voltage-current domain (VCD) and a manifold-dependent
one-dimensional map in the flux-charge domain (FCD), i.e., a one-dimensional map on each invariant manifold.
One main result is that, for suitable
choices of the circuit parameters and memristor nonlinearities, both DT circuits can exactly
embed two classic chaotic maps, i.e., the
\emph{logistic} map and the \emph{tent} map. Moreover, due to the property of extreme multistability,
the DT circuits can simultaneously embed in the manifolds all the dynamics displayed by
varying one parameter in the logistic and tent map. The paper then considers
a DT memristor Murali-Lakshmanan-Chua circuit and its dual. Via DT-FCAM these circuits implement a three-dimensional
map in the VCD and a two-dimensional map on each invariant manifold in the FCD.
It is shown that both circuits can simultaneously embed in the manifolds
all the dynamics displayed by two other classic chaotic maps, i.e., the \emph{H\'enon} map and the \emph{Lozi} map, when varying one parameter in such maps. In essence, these results provide
an explanation of why it is not surprising to observe complex dynamics even in
simple DT memristor circuits.
\end{abstract}


\section{Introduction}
\label{intro}

Memristor has been theoretically envisioned by Leon Chua in a seminal paper published in 1971 \cite{Chua1971} as the fourth basic passive circuit element in addition to resistor, capacitor and inductor. More than thirty years later, in 2007, a team at HP led by Stanley Williams made the fundamental discovery that we can physically observe memristive behavior in some metal-insulator-metal (MIM) solid-state devices \cite{Williams2008}.
This discovery has triggered an unprecedented and widespread interest in the theory and applications of continuous-time (CT) circuits with memristors, memcapacitors and meminductors \cite{spec-iss-memristor,mem_net,mem_Tetz,cfc2020,huang2021editorial,SpecIssue-InMemory2023,yang2013memristive,
sebastian2018tutorial,8470205,Sirakoulis2022717}. Among the many relevant potential applications there are the implementation of high-density and low power random access memories \cite{Zahoor2020,Sarwar201329,Hajri2019168963}. More generally, the memristor is seen as a strategic component for implementing in-memory computing schemes, akin to the computation mechanism of the brain, aiming at overcoming basic limitations of traditional Von Neumann computing machines \cite{williams2017s,zidan2018future,ielmini2020device,ascoli2022edgeSmale,kumar2017chaotic,liang2020s}.

Recent years have witnessed a significant surge of interest also for discrete time (DT) versions of memristor circuits. While the first DT circuits are obtained via Euler's discretization of simple CT memristor circuits \cite{bao2016extreme}, subsequent studies have also explored more abstract maps employing typical memristor nonlinearities but not necessarily related to the discretization of a CT counterpart \cite{ma2022locally,lai2023design,bao2022sine,9815581}. There are strong motivations for considering DT memristor circuits. First of all, since solid-state memristors are not easily available as yet, DT emulators can be useful to assess their performance without relying on experimental measurements. Moreover, DT memristor circuits can be easily and effectively interfaced with digital computers. However, the most relevant motivation is maybe that maps of DT memristor circuits can easily generate complex dynamics that are of interest in themselves for engineering applications. The main dynamical aspects studied in the literature include, but are not limited to, chaos and hyperchaos displayed by DT memristor maps \cite{peng2020discrete,li2020two,bao2021discrete,bao2022sine,9815581,ma2022locally,Deng20214601,Liu2023,
Liu20245094} and coexisting attractors, multistability and extreme multistability in memristor circuits with special classes of memristor nonlinearities \cite{bao2022parallel,lai2020two,lai2022hidden}. The generated complex dynamics have found relevant applications in the fields
of random number generation \cite{balatti2016physical,bao2021discrete}, secure communications \cite{li2020two,pisarchik2021secure}, implementation of reservoir computing systems \cite{deng20212d,Xu2023} and biomedical image encryption \cite{lin2022brain,Zhang202315487,Yang201958751}.

A recent line of research has addressed the extension to DT memristor circuits of the flux-charge analysis method (FCAM), a technique that was originally developed for CT memristor circuits \cite{Corinto-Forti-I,
Corinto-Forti-I,cfc2020}. In particular, the paper \cite{di2024new} has introduced a special discretization scheme ensuring that the first integrals of CT memristor circuits are exactly preserved in the discretization for any step size. As a consequence, the state space in the voltage-current domain (VCD) of DT memristor circuits can be foliated in invariant manifolds and on each manifold the circuits obey a manifold-dependent reduced-order dynamics in the flux-charge domain (FCD).
Such a structural property gives a rigorous explanation of
the so-called extreme multistability property of DT memristor circuits, i.e.,  the coexistence of infinitely many different dynamics and attractors for a fixed set of circuit parameters.

An inspection of the literature shows that even relatively simple DT memristor circuits can easily display really complex dynamics including chaos and hyperchaos \cite{bao2016extreme,di2024snap}. It is then natural to ask if we can find a rationale for this kind of behavior. Goal of this manuscript is to give an answer to this question via the study of the maps implemented via DT-FCAM by a simple DT memristor-capacitor circuit and a DT Murali-Lakshmanan-Chua circuit \cite{ishaq2013nonsmooth}. It is shown that, by a suitable choice of the memristor nonlinearity and circuit parameters, such circuits can embed, as a manifold-dependent reduced-order dynamics in the FCD, a number of classic chaotic maps, namely, the \emph{logistic and the tent map as well as the H\'enon and Lozi map \cite{hirsch2013differential,HK91}}. Another particularly intriguing feature is that by varying the manifold the \emph{memristor circuits can embed all the dynamics that are generated by varying one parameter in the logistic or H\'enon map.} This represents an effective way to exploit the property of coexisting attractors and extreme multistability, which is peculiar to DT memristor circuits \cite{di2024new}. In overall, the obtained results clearly show why it is not surprising to observe complex dynamics even in simple DT memristor circuits.

The organization of the manuscript can be outlined as follows. Sect.\ \ref{sect:DTFCAM} first
summarizes some results from DT-FCAM needed in the manuscript and then it derives the maps
implemented by the considered classes of DT memristor circuits. Thereafter, Sect.\ \ref{sect:embed logi}
and Sect.\ \ref{sect:embed henon} show how the logistic, tent, H\'enon and Lozi maps can be embedded in the DT memristor circuits. Finally, the main conclusions in the paper are given in
Sect.\ \ref{sect:concl}.

\section{Two Simple Classes of DT Memristor Circuits}
\label{sect:DTFCAM}
One basic question in dynamic system theory is whether it is possible to embed the dynamics of a given system into a higher-order system with a given structure. One relevant and quite unexpected result in this sense has been obtained by Stephen Smale in \cite{Sma76}. In that paper it is shown that any $n$-th order dynamics can be embedded in a competitive system of order $n+1$. Actually, the given dynamics evolves on an attracting $n$-th order invariant manifold of the competitive system. One problem however is that the equations of the embedding system do not resemble those of a network. An analogous result holds for cooperative systems, although in this case the $n$-th order manifold is a repelling one \cite{Hir89}. More recently, in \cite{Innocenti2022735} it has been shown how CT memristor circuits can embed the FitzHugh-Nagumo system
\cite{nagumo1962active} and the Duffing oscillators \cite{holmes1976bifurcations}.

In this manuscript, we address an embedding problem for a class of maps implemented by DT memristor circuits obtained via DT-FCAM \cite{di2024new}. More specifically, we study if it is possible to embed the logistic and the tent map, which are one-dimensional maps, as reduced-order dynamics in the invariant manifolds of a two-dimensional DT flux-controlled memristor and capacitor circuit or its dual (a charge-controlled memristor and inductor circuit). Moreover, we investigate whether we can embed the H\'enon and the Lozi map, which are two-dimensional maps, as reduced-order dynamics in the invariant manifolds of a three-dimensional DT Murali-Lakshmanan-Chua or its dual. First of all, in the remaining part of this section, we recall some basic facts on DT-FCAM useful for the analysis in the paper. Then, in Sect.\ \ref{sect:embed logi} and Sect.\ \ref{sect:embed henon}, we derive the maps implemented in the VCD and FCD by the considered classes of DT memristor circuits.

\subsection{Analysis via DT-FCAM}

FCAM has been developed in \cite{Corinto-Forti-I,Corinto-Forti-II} (see also the general treatment in \cite{cfc2020}) to analyze a class of CT circuits containing ideal resistors, capacitors, inductors and ideal flux-controlled or charge-controlled memristors. The method is based on analyzing the circuits both in the standard voltage current domain (VCD) and in the flux-charge domain (FCD) using Kirchhoff laws and constitutive relations of circuit elements. One relevant result obtained via FCAM is that for structural reasons a CT memristor circuit possesses first integrals in the VCD. Hence, the state space is foliated in invariant manifolds where the circuit satisfies a manifold-dependent reduced-order dynamics in the FCD. Later on, the recent article \cite{di2024new} has extended FCAM to DT memristor circuits. At the basis of DT-FCAM is the use of a special discretization scheme for the memristor circuits ensuring that the first integrals are exactly preserved in the discretization for any step size. This ensures that also a DT memristor circuit possesses integrals of motion and, as a consequence, its state space is foliated in invariant manifolds.

Consider a two-terminal element (a one-port) with voltage $v$ and current $i$. Let
$\varphi(t)=\int_{-\infty}^t v(\tau)d\tau$ and $q(t)=\int_{-\infty}^t i(\tau)d\tau$ be the flux and
charge, respectively. Moreover, consider the incremental
flux $\varphi^0(t)=\int_{0}^t v(\tau)d\tau$
and incremental charge $q^0(t)=\int_{0}^t i(\tau)d\tau$
\cite{Corinto-Forti-I}.
Given the sampling instants $t_k=kh$, $k=0,1,2,\dots$, where
$h>0$ is the step size, the corresponding discretized electric quantities are given by
$v_k=v(t_k)$, $i_k=i(t_k)$, $\varphi_k=\varphi(t_k)$, $q_k
=q(t_k)$, $\varphi_k^0=\varphi^0(t_k)$ and $q_k^0=q^0(t_k)$.

In the VCD, Kirchhoff voltage law (KVL) can be expressed as
$
\sum_j v_{j,k}=0
$
around any loop, whereas Kirchhoff current law (KCL) is given as
$
\sum_j i_{j,k}=0
$
for any cut-set.
In the FCD, Kirchhoff flux law (K$\varphi$L) is given as
$
\sum_j \varphi^0_{j,k}=0
$
around any loop, while Kirchhoff charge law (K$q$L) can be expressed as
$
\sum_j q^0_{j,k}=0
$
for any cut-set \cite{di2024new}.

Consider a linear capacitor $q_C=Cv_C$ or $i_C=Cdv_C/dt$.
Via forward Euler rule we obtain the one-dimensional map (VCD)
\begin{equation}\label{C-DT-CVD}
    i_{C,k}=C\frac{v_{C,{k+1}}-v_{C,k}}{h}.
\end{equation}
Moreover, in the FCD the capacitor satisfies the one-dimensional map
\begin{equation}\label{C-DT-FCD}
    q^0_{C,k}=C  \frac{\varphi_{C,k+1}^0-\varphi_{C,k}^0}{h}-Cv_{C0}
\end{equation}
where $v_{C0}=v_C(0)$.

Consider now a linear inductor $\varphi_L=Li_L$ or $v_L=Ldi_L/dt$.
Applying forward Euler rule we obtain the one-dimensional map (VCD)
\begin{equation}\label{L-DT-CVD}
    v_{L,k}=L\frac{i_{L,{k+1}}-i_{L,k}}{h}.
\end{equation}
Moreover, in the FCD the inductor satisfies the one-dimensional map
\begin{equation}\label{L-DT-FCD}
    \varphi^0_{L,k}=L  \frac{q_{L,k+1}^0-q_{L,k}^0}{h}-Li_{L0}
\end{equation}
where $i_{L0}=i_L(0)$

A flux-controlled memristor is defined by $q_M=\hat q (\varphi_M)$.
This yields (VCD)
\begin{equation}\label{M-CT-VCD}
\left\{
  \begin{array}{ll}
    i_M=\hat q'(\varphi_M)v_M \\
    \dot \varphi_M=v_M.
  \end{array}
\right.
\end{equation}
In accordance with \cite{di2024new}, we use for the memristor the discretization scheme (VCD)
\begin{equation}\label{S2 Mp-DT-FCD}
    \left\{
  \begin{array}{ll}
    i_{M,k}=\frac{\hat q(\varphi_{M,{k+1}})
-\hat q(\varphi_{M,k})}{h} \\
    \varphi_{M,{k+1}}=\varphi_{M,k}+hv_{M,k}.
  \end{array}
\right.
\end{equation}
On the other hand we have (FCD)
\begin{equation}\label{M-DT-FCD}
    q^0_{M,k}=\hat q(\varphi_{M,k}^0+\varphi_{M0})-\hat q(\varphi_{M0})
\end{equation}
where $\varphi_{M0}=\varphi_M(0)$.

A charge-controlled memristor is defined by $\varphi_M=\hat \varphi (q_M)$.
This yields (VCD)
\begin{equation}\label{Mi-CT-VCD}
\left\{
  \begin{array}{ll}
    v_M=\hat \varphi'(q_M)i_M \\
    \dot q_M=i_M.
  \end{array}
\right.
\end{equation}
In accordance with \cite{di2024new}, we use for the memristor the discretization scheme (VCD)
\begin{equation}\label{S2 Mi-DT-FCD}
    \left\{
  \begin{array}{ll}
    v_{M,k}=\frac{\hat \varphi(q_{M,{k+1}})
-\hat \varphi(q_{M,k})}{h} \\
    q_{M,{k+1}}=q_{M,k}+hi_{M,k}.
  \end{array}
\right.
\end{equation}
On the other hand we have (FCD)
\begin{equation}\label{Mi-DT-FCD}
    \varphi^0_{M,k}=\hat \varphi(q_{M,k}^0+q_{M}(0))-\hat \varphi(q_{M0})
\end{equation}
where $\varphi_{M0}=\varphi_M(0)$.

\begin{remark}
We stress that the discretization schemes for the memristor given in (\ref{S2 Mp-DT-FCD})
and (\ref{S2 Mi-DT-FCD}) guarantee that the first integrals of CT memristor circuits are
preserved in the discretization for any step size \cite{di2024new}. As such, they
basically differ from typical discretization schemes used in the literature that do not
preserve first integrals \cite{PhysDhenon}. \qed
\end{remark}

\subsection{MC and ML circuits}

\begin{figure}[ht]
\begin{center}
\includegraphics[width=0.5\columnwidth]{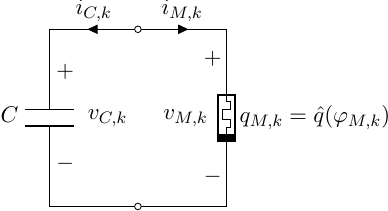} \\
(a) ~\\
~\\
\includegraphics[width=0.5\columnwidth]{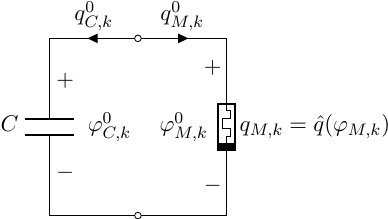} \\
(b)
\end{center}
\caption{DT MC circuit. (a) VCD and (b) FCD.}
\label{fig:MC_circuit}
\end{figure}

First, let us consider the MC circuit in Fig.~\ref{fig:MC_circuit}(a)
(VCD), which is composed of an ideal flux-controlled
memristor and an ideal capacitor. The corresponding circuit in the FCD is
shown in Fig.~\ref{fig:MC_circuit}(b). The next result provides us with
the maps implemented by the circuit in the VCD and FCD. It also gives the
expression of the first integral and the invariant manifolds of the circuit.

\begin{prop}
\label{map-MC}
The following results hold:

1) In the VCD the dynamics of the MC circuit is described by the two-dimensional
map in the state variables
$v_{C,k}$ and $\varphi_{M,k}$
\begin{equation}\label{MCirc-VCD}
\left\{
  \begin{array}{ll}
    v_{C,k+1}=v_{C,k}-\frac{1}{C}(\hat q(\varphi_{M,k}+hv_{C,k})-\hat
q(\varphi_{M,k})) \\
    \varphi_{M,k+1}=\varphi_{M,k}+hv_{C,k}.
  \end{array}
\right.
\end{equation}

2) The function of the state variables given by
$$
\Theta_\mathrm{MC}(v_C,\varphi_M)= Cv_C+\hat q(\varphi_M)
$$
is a first integral of (\ref{MCirc-VCD}) for any step size $h$, i.e., we have
$
\Theta_\mathrm{\mathrm{MC}}(v_{C,k+1},\varphi_{M,k+1})=\Theta_\mathrm{MC}(v_{C,k},\varphi_{M,k}),
$
$k=0,1,2,\dots$.

3) The state space of (\ref{MCirc-VCD})
is foliated in one-dimensional invariant manifolds
\begin{align*}
\cM_\mathrm{MC}(Q_0) &= \{ (v_C,\varphi_M) \in \R^2:\\
 & \Theta_\mathrm{MC}(v_C,\varphi_M)=Cv_C+\hat q (\varphi_M)=Q_0 \}
\end{align*}
where $Q_0 \in \R$ is the manifold index. On each manifold, i.e.,
in the FCD, the dynamics
is described by the one-dimensional map in the state variable $\varphi_{M,k}$
\begin{equation}\label{MC map FC}
    \varphi_{M,k+1}=\varphi_{M,k}-\frac{h}{C}\hat q(\varphi_{M,k})
+\frac{h}{C}Q_0.
\end{equation}
\qed
\end{prop}

{\em Proof}. 1) KCL and KVL written for the MC circuit in
Fig.~\ref{fig:MC_circuit}(a) yield
\begin{align}
i_{C,k}+ i_{M,k}&=0 \label{MC_EQ_KI} ,\\
v_{C,k}&=v_{M,k} \label{MC_EQ_KV}.
\end{align}
Substituting~(\ref{C-DT-CVD}) and the first equation of~(\ref{S2
Mp-DT-FCD}) in~(\ref{MC_EQ_KI}) we obtain
\begin{equation}
\label{MC_State_VCD1}
C \frac{v_{C,k+1}-v_{C,k}}{h} + \frac{\hat{q}(\varphi_{M k+1})-
\hat{q}(\varphi_{M k})}{h}=0.
\end{equation}

Equation~(\ref{MC_EQ_KV}) allows us to write the second equation
of~(\ref{S2 Mp-DT-FCD}) as
\begin{equation}
\label{MC_State_VCD2}
\varphi_{M,k+1}=\varphi_{M,k} + h v_{C,k}.
\end{equation}
Now, (\ref{MC_State_VCD1}) and~(\ref{MC_State_VCD2}) can be
straightforwardly recast into~(\ref{MCirc-VCD}).

2) Follows directly by reordering the terms of~(\ref{MC_State_VCD1}).

3) From~(\ref{MC_State_VCD1}) we have that that, for any $k \ge 1$,
$$
Cv_{C,k}+\hat q (\varphi_{M,k})=Cv_{C,0}+\hat q (\varphi_{M,0})=Q_0
$$
i.e., the state variables are bound to evolve on the manifold
$\cM_\mathrm{MC}(Q_0)$. To derive the equations describing the dynamics on $\cM_\mathrm{MC}(Q_0)$,
let us write the K$q$L and K$\varphi$L for the MC circuit in
Fig.~\ref{fig:MC_circuit}(b)
\begin{align}
q^0_{C,k}+ q^0_{M,k}&=0 \label{MC_EQ_KQ}\\
\varphi^0_{C,k}&=\varphi^0_{M,k} \label{MC_EQ_KF}.
\end{align}
Substituting~(\ref{C-DT-FCD}) and~(\ref{M-DT-FCD}) in~(\ref{MC_EQ_KQ}) we
obtain
\begin{align*}
C & \frac{\varphi^0_{C,k+1}-\varphi^0_{C,k}}{h} - C v_{C,0} \\
& + \hat{q}(\varphi^0_{M,k}+\varphi_{M,0}) - \hat{q}(\varphi_{M,0})=0.
\end{align*}
Considering~(\ref{MC_EQ_KF}), we obtain
$$
\varphi^0_{M,k+1}= \varphi^0_{M,k} - \frac{h}{C}
\hat{q}(\varphi^0_{M,k}+ \varphi_{M,0}) + h v_{C,0} +
\frac{h}{C}\hat{q}(\varphi_{M,0})
$$
which can be recast into~(\ref{MC map FC}) by summing $\varphi_{M,0}$ to
both sides and taking into account the expression of $Q_0$. \qed \\

\begin{figure}[ht]
\begin{center}
\includegraphics[width=0.5\columnwidth]{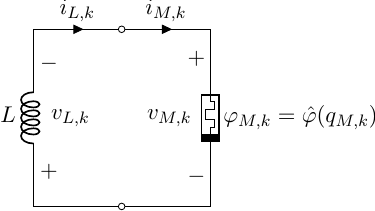} \\
(a) ~\\
~\\
\includegraphics[width=0.5\columnwidth]{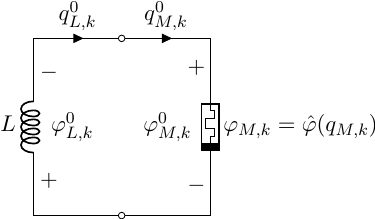} \\
(b)
\end{center}
\caption{DT ML circuit. (a) VCD and (b) FCD}
\label{fig:ML_circuit}
\end{figure}

Now, let us consider the dual of the circuit in Fig.~\ref{fig:MC_circuit},
which is composed of an ideal charge-controlled
memristor and an ideal inductor, as depicted in
Fig.~\ref{fig:ML_circuit}(a) (VCD). The corresponding circuit in the FCD
is shown in Fig.~\ref{fig:ML_circuit}(b).

\begin{prop}
\label{map-ML}
The following results hold:

1) In the VCD the dynamics of the ML circuit is described by the two-dimensional
map in the state variables
$i_{L,k}$ and $q_{M,k}$
\begin{equation}\label{MLCirc-VCD}
\left\{
  \begin{array}{ll}
    i_{L,k+1}=i_{L,k}-\frac{1}{L}(\hat \varphi(q_{M,k}+hi_{L,k})-\hat
\varphi(q_{M,k})) \\
    q_{M,k+1}=q_{M,k}+hi_{L,k}.
  \end{array}
\right.
\end{equation}

2) The function of the state variables given by
$$
\Theta_\mathrm{ML}(i_L,q_M)= Li_L+\hat \varphi(q_M)
$$
is a first integral of (\ref{MLCirc-VCD}) for any step size $h$, i.e., we
have
$
\Theta_\mathrm{\mathrm{ML}}(i_{L,k+1},q_{M,k+1})=\Theta_\mathrm{ML}(i_{L,k},q_{M,k}),
$
$k=0,1,2,\dots$.

3) The state space of (\ref{MLCirc-VCD})
is foliated in one-dimensional invariant manifolds
\begin{align*}
\cM_\mathrm{ML}(\Phi_0) &= \{ (i_L,q_M) \in \R^2:\\
 & \Theta_\mathrm{ML}(i_L,q_M)=Li_L+\hat \varphi (q_M)=\Phi_0 \}
\end{align*}
where $\Phi_0 \in \R$ is the manifold index. On each manifold, i.e.,
in the FCD, the dynamics
is described by the one-dimensional map in the state variable $q_{M,k}$
\begin{equation}\label{ML map FC}
    q_{M,k+1}=q_{M,k}-\frac{h}{L}\hat \varphi(q_{M,k})
+\frac{h}{L} \Phi_0.
\end{equation}
\qed
\end{prop}

{\em Proof}. 1) KCL and KVL for the ML circuit in
Fig.~\ref{fig:ML_circuit}(a) are given as
\begin{align}
v_{L,k}+ v_{M,k}&=0 \label{ML_EQ_KV}\\
i_{L,k}&=i_{M,k} \label{ML_EQ_KI}.
\end{align}
Substituting~(\ref{L-DT-CVD}) and the first equation of~(\ref{S2
Mi-DT-FCD}) in~(\ref{ML_EQ_KV}) we obtain
\begin{equation}
\label{ML_State_VCD1}
L \frac{i_{L,k+1}-i_{L,k}}{h} + \frac{\hat{\varphi}(q_{M k+1})-
\hat{\varphi}(q_{M k})}{h}=0.
\end{equation}

Equation~(\ref{ML_EQ_KI}) allows us to write the second equation
of~(\ref{S2 Mi-DT-FCD}) as
\begin{equation}
\label{ML_State_VCD2}
q_{M,k+1}=q_{M,k} + h i_{L,k}.
\end{equation}
Now, it is seen that~(\ref{ML_State_VCD1}) and~(\ref{ML_State_VCD2}) can be
straightforwardly recast into~(\ref{MLCirc-VCD}).

2) Follows directly by reordering the terms of~(\ref{ML_State_VCD1}).

3) From~(\ref{ML_State_VCD1}) we have that, for any $k>1$,
$$
Li_{L,k}+\hat \varphi (q_{M,k})=Li_{L,0}+\hat \varphi (q_{M,0})=\Phi_0
$$
i.e., the state variables are bound to evolve on the manifold
$\cM_\mathrm{ML}(\Phi_0)$. To derive the equations describing
the dynamics on such a manifold, let
us first write the K$q$L and K$\varphi$L for the ML circuit in
Fig.~\ref{fig:ML_circuit}(b)
\begin{align}
q^0_{L,k}&= q^0_{M,k} \label{ML_EQ_KQ}\\
\varphi^0_{L,k}+\varphi^0_{M,k}&=0 \label{ML_EQ_KF}.
\end{align}
Substituting~(\ref{L-DT-FCD}) and~(\ref{Mi-DT-FCD}) in~(\ref{ML_EQ_KF}) we
obtain
\begin{align*}
L & \frac{q^0_{L,k+1}-q^0_{L,k}}{h} - L i_{L,0} \\
& + \hat{\varphi}(q^0_{M,k}+q_{M,0}) - \hat{\varphi}(q_{M,0})=0.
\end{align*}
Considering~(\ref{ML_EQ_KQ}), we have
$$
q^0_{M,k+1}= q^0_{M,k} - \frac{h}{L} \hat{\varphi}(q^0_{M,k}+ q_{M,0}) +
h i_{L,0} + \frac{h}{L}\hat{\varphi}(q_{M,0})
$$
which can be recast into~(\ref{ML map FC}) by summing $q_{M,0}$ to both
sides and taking into account the expression of $\Phi_0$. \qed \\

Clearly, via a suitable change of variables, it is possible to put the maps,
first integral and invariant manifolds of the MC circuit and its dual
in the same abstract mathematical form. Indeed, for the MC circuit, let
$x_k=\varphi_{M,k}$, $y_k=v_{C,k}$, $\alpha=1/C$, $F(\cdot)=\hat q(\cdot)$
and $I_0=Q_0$. Moreover, for the ML circuit, let $x_k=q_{M,k}$,
$y_k=i_{L,k}$,
$\alpha=1/L$, $F(\cdot)=\hat \varphi(\cdot)$ and $I_0=\Phi_0$. Then,
for both circuits the dynamics in the VCD obeys the
two-dimensional map
in the state variables $x_k$ and $y_k$
\begin{equation}\label{map-VCD}
\left\{
  \begin{array}{ll}
    y_{k+1}=y_k-\alpha(F(x_k+hy_k)-F(x_k)) \\
    x_{k+1}=x_k+hy_k.
  \end{array}
\right.
\end{equation}
Moreover, there exists a first integral given by
\begin{equation}\label{integral MCML}
    \Theta(x,y)= \frac{1}{\alpha}y+F(x)
\end{equation}
and the state space in the VCD is foliated in the invariant manifolds
\begin{align*}
\cM(I_0)= \{ (x,y) \in \R^2: \Theta(x,y)=\frac{1}{\alpha}y+F(x)=I_0 \}
\end{align*}
where $I_0 \in \R$ is the manifold index.
On each manifold the dynamics is described in the FCD by the one-dimensional map
in the state variable $x_k$
\begin{equation}\label{map FC}
    x_{k+1}=x_k-h\alpha F(x_k)+h \alpha I_0.
\end{equation}

\subsection{Memristor Murali-Lakshmanan-Chua circuit and its dual}
\begin{figure}[ht]
\begin{center}
\includegraphics[width=0.7\columnwidth]{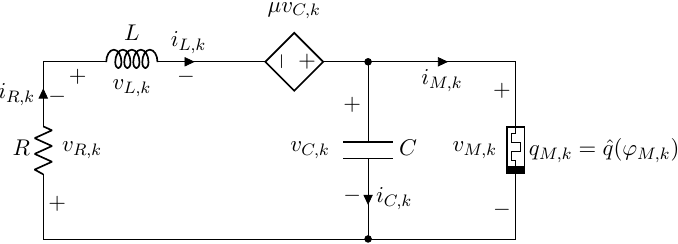} \\
(a) ~\\
~\\
\includegraphics[width=0.7\columnwidth]{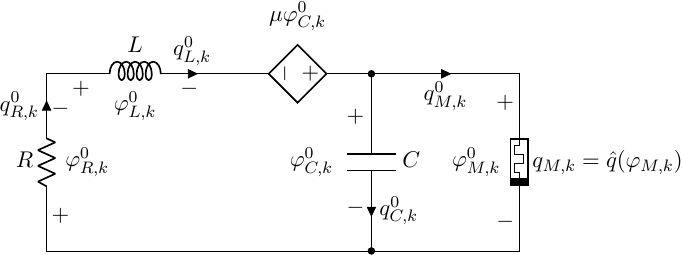} \\
(b)
\end{center}
\caption{DT MLC circuit. (a) VCD and (b) FCD.}
\label{fig:MLC_circuit}
\end{figure}

Fig.~\ref{fig:MLC_circuit}(a) depicts a circuit in the VCD,
obtained from Murali-Lakshmanan-Chua (MLC) circuit
by replacing the nonlinear resistor with a flux-controlled
memristor \cite{murali1994simplest,ishaq2013nonsmooth,Itoh2008,Innocenti2022735}. For more
generality, a voltage-controlled voltage-source (VCVS) is also added in series
with the inductor. The corresponding circuit in the FCD is shown in Fig.~\ref{fig:MLC_circuit}(b).
The MLC circuit is a simplified version of the celebrated Chua's circuit \cite{procIEEE-chaos}.

\begin{prop}
\label{map-MLC}
The following results hold:

1) In the VCD the dynamics of MLC circuit is described by the three-dimensional map in
the state variables
$v_{C,k}, i_{L,k}$ and $\varphi_{M,k}$
\begin{equation}\label{MLC-VCD}
   \left\{
  \begin{array}{ll}
     v_{C,k+1}=v_{C,k}-\frac{\hat q(\varphi_{M,k}+hv_{C,k})-\hat
q(\varphi_{M,k})}{C} + \frac{h}{C}i_{L,k}\\
     i_{L,k+1}= (1-\frac{Rh}{L}) i_{L,k}+\frac{h}{L}(\mu -1) v_{C,k}\\
    \varphi_{M,k+1}=\varphi_{M,k}+hv_{C,k}.
  \end{array}
\right.
\end{equation}

2) The function of the state variables given by
$$
\Theta_\mathrm{MLC}(v_C,i_L,\varphi_M)= Cv_C+\frac{Li_L}{R}+\hat
q(\varphi_M)+\frac{1-\mu}{R}\varphi_M
$$
is a first integral of (\ref{MLC-VCD}) for any step size $h$, i.e., we have
$
\Theta_\mathrm{\mathrm{MLC}}(v_{C,k+1},i_{L,k+1},\varphi_{M,k+1})=
\Theta_\mathrm{MLC}(v_{C,k},i_{L,k},\varphi_{M,k}),
$
$k=0,1,2,\dots$.

3) The state space of (\ref{MLC-VCD})
is foliated in two-dimensional invariant manifolds
\begin{align*}
\cM_\mathrm{MLC}(Q_0) &= \{ (v_C,i_L,\varphi_M) \in \R^3:
\Theta_\mathrm{MLC}(v_C,i_L,\varphi_M)\\
&=Cv_C+\frac{Li_L}{R}+\hat q(\varphi_M)
 +\frac{1-\mu}{R}\varphi_M=Q_0 \}
\end{align*}
where $Q_0 \in \R$ is the manifold index. On each manifold, i.e., in the
FCD, the dynamics
is described by the two-dimensional map in the state variables $\varphi_{M,k}$
and $y_k=q^0_{L,k}+\frac{\mu-1}{R}\varphi_{M,0}-\frac{L}{R}i_{L,0}$
\begin{equation}\label{MLC FCD}
    \left\{
  \begin{array}{lcl}
     \varphi_{M,k+1}&=&\varphi_{M,k}+ \frac{h}{C}y_k-\frac{h}{C}\hat
q(\varphi_{M,k})+\frac{h}{C}Q_0\\
     y_{k+1}&=& (1-\frac{hR}{L}) y_k+\frac{h}{L}(\mu-1)\varphi_{M,k}.
  \end{array}
\right.
\end{equation}
\qed
\end{prop}

{\em Proof}.
1) First, we can write KCLs and KVLs for the MLC circuit in
Fig.~\ref{fig:MLC_circuit}(a) as follows
\begin{align}
i_{L,k}&=i_{M,k}+i_{C,k} \label{MLC_KI_LMC}\\
i_{L,k}&=i_{R,k} \label{MLC_KI_LR}\\
v_{C,k}+v_{R,k} + v_{L,k}-\mu v_{C,k} &=0 \label{MLC_KV_CLR}\\
v_{C,k}&= v_{M,k}. \label{MLC_KV_CM}
\end{align}
Substituting~(\ref{C-DT-CVD}) and the first equation of~(\ref{S2
Mp-DT-FCD}) in~(\ref{MLC_KI_LMC}) we obtain
\begin{equation}
\label{MCL_State_VCD1}
i_{L,k}=C \frac{v_{C,k+1}-v_{C,k}}{h}+ \frac{\hat q(\varphi_{M, k+1})-
\hat q( \varphi_{M,k})}{h}
\end{equation}
while using~(\ref{L-DT-CVD}) and~(\ref{MLC_KI_LR}),
from~(\ref{MLC_KV_CLR}) we have
\begin{equation}
\label{MCL_State_VCD2}
(1-\mu) v_{C,k}+R i_{L,k} + L \frac{i_{L,k+1}-i_{L,k}}{h}=0.
\end{equation}
Using~(\ref{MLC_KV_CM}) we can write the second equation
of~(\ref{S2 Mp-DT-FCD}) as
\begin{equation}
\label{MCL_State_VCD3}
\varphi_{M,k+1}=\varphi_{M,k} + h v_{C,k}.
\end{equation}
Now, (\ref{MCL_State_VCD1})-(\ref{MCL_State_VCD3}) can be
straightforwardly recast into~(\ref{MLC-VCD}).

2) Solving~(\ref{MCL_State_VCD3}) with respect to $v_{C,k}$ and substituting
in~(\ref{MCL_State_VCD2}) we obtain the following expression
\begin{equation*}
i_{L,k}= (\mu-1)
\frac{\varphi_{M,k+1}-\varphi_{M,k}}{Rh}-L\frac{i_{L,k+1}-i_{L,k}}{Rh}.
\end{equation*}
By using~(\ref{MCL_State_VCD1}) and reordering terms we arrive at
\begin{align}
C& v_{C,k+1}+ \frac{L i_{L,k+1}}{R} + \hat{q}(\varphi_{M,k+1})+
\frac{1-\mu}{R} \varphi_{M, k+1} \label{MANI-MLC}\\
&= C v_{C,k}+ \frac{L i_{L,k}}{R} + \hat{q}(\varphi_{M,k})+
\frac{1-\mu}{R}\varphi_{M, k}. \nonumber
\end{align}

3) From~(\ref{MANI-MLC}) we have that, for any $k>1$,
$$
C v_{C,k}+ \frac{L i_{L,k}}{R} + \hat{q}(\varphi_{M,k})+ \frac{1-\mu}{R} \varphi_{M,
k}= Q_0
$$
i.e., the state variables are bound to evolve on the manifold
$\cM_\mathrm{MLC}(Q_0)$.
To derive the equations describing the dynamics on such a manifold, let us write the K$q$L and
K$\varphi$L for the MLC circuit in Fig.~\ref{fig:MLC_circuit}(b)
\begin{align}
q^0_{L,k}&=q^0_{M,k}+q^0_{C,k}, \label{MLC_KQ_LMC}\\
\varphi^0_{C,k}+\varphi^0_{R,k}+\varphi^0_{L,k} - \mu \varphi^0_{C,k} &=0 \label{MLC_KF_CLR}\\
\varphi^0_{C,k}&= \varphi^0_{M,k}. \label{MLC_KF_CM}
\end{align}
Substituting~(\ref{C-DT-FCD}) and~(\ref{M-DT-FCD}) in
(\ref{MLC_KQ_LMC}) and~(\ref{L-DT-FCD}) in (\ref{MLC_KF_CLR}), and
taking into account~(\ref{MLC_KF_CM}) we have
\begin{align*}
\varphi^0_{M,k+1}&=\varphi^0_{M,k}- \frac{h}{C}\hat q (\varphi^0_{M,k} +
\varphi_{M,0})+ \frac{h}{C} q^0_{L,k} \\
& + h v_{C,0}+ \frac{h}{C} \hat q (\varphi_{M,0}) \\
q^0_{L,k+1} & = \left( 1- \frac{hR}{L} \right) q^0_{L,k} + h i_{L,0} +
\frac{h}{L} (\mu -1) \varphi^0_{M,k}
\end{align*}
which can be recast into (\ref{MLC FCD}) by summing $\varphi_{M,0}$ to both
sides of the first equation and taking into account the expressions of
$y_k$ and $Q_0$.
\qed

\begin{figure}[ht]
\begin{center}
\includegraphics[width=0.7\columnwidth]{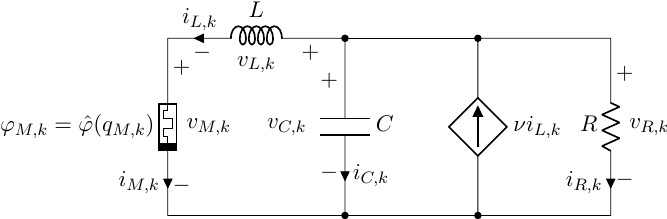} \\
(a) ~\\
~\\
\includegraphics[width=0.7\columnwidth]{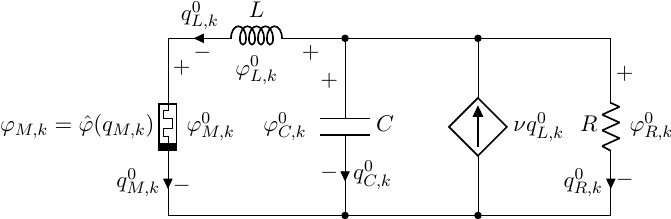} \\
(b)
\end{center}
\caption{DT DMLC circuit. (a) VCD and (b) FCD}
\label{fig:MLC_dual_circuit}
\end{figure}

Finally, let us consider the dual of the circuit in Fig.~\ref{fig:MLC_circuit},
i.e., DMLC, which is depicted in Fig.~\ref{fig:MLC_dual_circuit}(a) (VCD). The
corresponding circuit in the FCD is shown in
Fig.~\ref{fig:MLC_dual_circuit}(b).

\begin{prop}
\label{map-MLC-dual}
The following results hold:

1) In the VCD the dynamics of DMLC is described by the
three-dimensional map in the state variables
$i_{L,k}, v_{C,k}$ and $q_{M,k}$
\begin{equation}\label{DMLC-VCD}
   \left\{
  \begin{array}{ll}
     i_{L,k+1}=i_{L,k}-\frac{\hat \varphi(q_{M,k}+hi_{L,k})-\hat
\varphi(q_{M,k})}{L} + \frac{h}{L}v_{C,k}\\
     v_{C,k+1}= (1-\frac{h}{RC}) v_{C,k}+\frac{h}{C}(\nu -1) i_{L,k}\\
    q_{M,k+1}=q_{M,k}+hi_{L,k}.
  \end{array}
\right.
\end{equation}

2) The function of the state variables given by
$$
\Theta_\mathrm{DMLC}(v_C,i_L,q_M)= RCv_C+Li_L+\hat \varphi(q_M)+R(1-\nu)q_M
$$
is a first integral of (\ref{MLC-VCD}) for any step size $h$, i.e., we have
$
\Theta_\mathrm{\mathrm{DMLC}}(v_{C,k+1},i_{L,k+1},q_{M,k+1})=
\Theta_\mathrm{DMLC}(v_{C,k},i_{L,k},q_{M,k}),
$
$k=0,1,2,\dots$.

3) The state space of (\ref{MLC-VCD})
is foliated in two-dimensional invariant manifolds
\begin{align*}
\cM_\mathrm{DMLC}(\Phi_0) &= \{ (v_C,i_L,q_M) \in \R^3:
\Theta_\mathrm{DMLC}(v_C,i_L,q_M)\\
&=CRv_C+Li_L+\hat \varphi(q_M) + R (1-\nu) q_M\\
&=\Phi_0 \}
\end{align*}
where $\Phi_0 \in \R$ is the manifold index. On each manifold, i.e., in the
FCD, the dynamics
is described by the two-dimensional map in the state variables $q_{M,k}$
and $y_k=\varphi^0_{C,k}+R(\nu-1)q_{M,0}-CRv_{C,0}$
\begin{equation}\label{DMLC-FCD}
    \left\{
  \begin{array}{lcl}
     q_{M,k+1}&=&q_{M,k}+ \frac{h}{L}y_k-\frac{h}{L}\hat
\varphi(q_{M,k})+\frac{h}{L} \Phi_0\\
     y_{k+1}&=& (1-\frac{h}{RC}) y_k+\frac{h}{C} (\nu -1) q_{M,k}.
  \end{array}
\right.
\end{equation}
\qed
\end{prop}

{\em Proof}. 1) First, write KCLs and KVLs for the DMLC circuit in
Fig.~\ref{fig:MLC_dual_circuit}(a) as follows
\begin{align}
v_{C,k}&=v_{L,k}+v_{M,k}  \label{DMLC_KV_CLM}\\
v_{C,k}&= v_{R,k} \label{DMLC_KV_CR}\\
i_{L,k}+i_{R,k}+i_{C,k} - \nu i_{L,k} &=0 \label{DMLC_KI_LCR}\\
i_{L,k}&=i_{M,k}. \label{DMLC_KI_LM}
\end{align}
Substituting~(\ref{L-DT-CVD}) and the first equation~(\ref{S2 Mi-DT-FCD})
in~(\ref{DMLC_KV_CLM}) we obtain
\begin{equation}
\label{DMCL_State_VCD1}
v_{C_k}=L \frac{i_{L,k+1}-i_{L,k}}{h}+ \frac{\hat \varphi(q_{M, k+1})-
\hat \varphi(q_{M,k})}{h}
\end{equation}
while using~(\ref{C-DT-CVD}) and~(\ref{DMLC_KV_CR}),
from~(\ref{DMLC_KI_LCR}) we obtain
\begin{equation}
\label{DMCL_State_VCD2}
(1-\nu)i_{L,k}+C \frac{v_{C,k+1}-v_{C,k}}{h}+ \frac{v_{C,k}}{R}=0.
\end{equation}
Exploiting~(\ref{DMLC_KI_LM}) we can write the second equation
of~(\ref{S2 Mi-DT-FCD}) as
\begin{equation}
\label{DMCL_State_VCD3}
q_{M,k+1}=q_{M,k} + h i_{L,k}.
\end{equation}
Now, ~(\ref{DMCL_State_VCD1})-(\ref{DMCL_State_VCD3}) can be
straightforwardly recast into~(\ref{DMLC-VCD}).

2) Solving~(\ref{DMCL_State_VCD3}) with respect to $i_{L,k}$ and substituting
in~(\ref{DMCL_State_VCD2}) we obtain the following expression
\begin{equation*}
v_{C,k}= R (\nu -1)\frac{q_{M,k+1}-q_{M,k}}{h}-RC\frac{v_{C,k+1}-v_{C,k+1}}{h}.
\end{equation*}
Equating the right hand sides of the latter equation
and~(\ref{DMCL_State_VCD1}) and reordering the terms we have
\begin{align}
RC& v_{C,k+1} + L i_{L,k+1} + \hat{\varphi}(q_{M,k+1})+R(1-\nu)q_{M, k+1}
\label{MANI-DMLC}\\
&=RC v_{C,k}+ L i_{L,k} + \hat{\varphi}(q_{M,k})+R(1-\nu)q_{M, k} . \nonumber
\end{align}

3) Using~(\ref{MANI-DMLC}) we have that, for any $k>1$,
$$
RC v_{C,k}+ L i_{L,k} + \hat{\varphi}(q_{M,k})+R(1-\nu)q_{M, k} = \Phi_0
$$
i.e., the state variables are bound to evolve on the manifold
$\cM_\mathrm{DMLC}(\Phi_0)$.
To write the dynamics on such a manifold, let us write the K$q$L and
K$\varphi$L for the DMLC circuit in Fig.~\ref{fig:MLC_dual_circuit}(b)
\begin{align}
\varphi^0_{C,k}&=\varphi^0_{M,k}+\varphi^0_{L,k} \label{DMLC_KF_LMC}\\
q^0_{C,k}+q^0_{L,k}+q^0_{R,k} - \nu q^0_{L,k} &=0 \label{DMLC_KQ_CLR}\\
q^0_{L,k}&= q^0_{M,k}. \label{DMLC_KQ_LM}
\end{align}
Substituting~(\ref{L-DT-FCD}) and~(\ref{Mi-DT-FCD}) in
(\ref{DMLC_KF_LMC}) and~(\ref{C-DT-FCD}) in (\ref{DMLC_KQ_CLR}),
and taking into account~(\ref{DMLC_KQ_LM}) we obtain
\begin{align*}
q^0_{M,k+1}&=q^0_{M,k}- \frac{h}{L}\hat \varphi (q^0_{M,k} + q_{M,0})+
\frac{h}{L} \varphi^0_{L,k} \\
& + h i_{L,0}+ \frac{h}{L} \hat \varphi (q_{M,0}) \\
\varphi^0_{C,k+1} & = \left(1- \frac{h}{RC}\right) \varphi^0_{C,k} + h v_{C,0} +
\frac{h}{C} (\nu -1) q^0_{M,k}.
\end{align*}
This can be recast as~(\ref{DMLC-FCD}) by summing $q_{M,0}$ to both sides
of the first equation, and taking into account the expressions of $y_k$
and $\Phi_0$.
\qed \

For the MLC, let $x_k=\varphi_{M,k}$, $z_k=i_{L,k}$, $w_k=v_{C,k}$,
$F(\cdot)=
\hat q(\cdot)$, $\alpha=1/C$, $\beta=R/L$, $\gamma=(\mu-1)/L$, while for DMLC, let
$x_k=q_{M,k}$, $z_k=v_{C,k}$, $w_k=i_{L,k}$, $F(\cdot)=
\hat \varphi(\cdot)$, $\alpha=1/L$, $\beta=1/(CR)$, $\gamma=(\nu-1)/C$.
Then, the dynamics in the VCD of both circuits obeys the
three-dimensional map
in the state variables $w_k,z_k$ and $x_k$
\begin{equation}\label{bothMLC-VCD}
\left\{
  \begin{array}{ll}
    w_{k+1}=w_k-\alpha(F(x_k+hw_k)-F(x_k))+\alpha h z_k \\
    z_{k+1}=(1-h\beta)z_k+h\gamma w_k\\
    x_{k+1}=x_k+h w_k.
  \end{array}
\right.
\end{equation}

There is a first integral given by
\begin{equation}\label{integral MLC}
\Theta(x,w,z)= \frac{1}{\alpha}w+\frac{1}{\beta}z-\frac{\gamma}{\beta}x+F(x)
\end{equation}
and the state space in the VCD is foliated in the invariant manifolds
\begin{align*}
\cM(I_0) &= \{ (x,y,z) \in \R^3: \Theta(x,y,z)=
\frac{1}{\alpha}w+\frac{1}{\beta}z\\
 &-\frac{\gamma}{\beta}x+F(x)=   I_0 \}
\end{align*}
where $I_0 \in \R$ is the manifold index.

Moreover, let for MLC $y_k=q^0_{L,k}-\frac{z_0}{\beta}+\frac{\gamma}{\beta}x_0$ and
$I_0=Q_0$, while for DMLC,
let $y_k=\varphi^0_{C,k}-\frac{z_0}{\beta}+\frac{\gamma}{\beta}x_0$ and $I_0=\Phi_0$. Then,
on each invariant manifold the dynamics in the FCD of both
circuits is
described by the two-dimensional map in the state variables $x_k$ and $y_k$
\begin{equation}\label{bothMLC-FCD}
\left\{
  \begin{array}{ll}
    x_{k+1}=x_k +h\alpha y_k-h\alpha F(x_k)+h\alpha I_0 \\
    y_{k+1}=y_k-h\beta y_k+h\gamma x_k.
  \end{array}
\right.
\end{equation}

\section{Embedding the Logistic and Tent Map in the MC and ML Circuit}
\label{sect:embed logi}
\subsection{Logistic map}

The logistic map became popular due to a paper published in Nature in 1976
by biologist Robert May
and entitled `Simple mathematical models with very complicated dynamics'
\cite{may1976simple}. The logistic map is an equation for population growth
\begin{equation}\label{logistic}
    P_{k+1}=rP_k(1-P_k)
\end{equation}
where the next generation population
$P_{k+1}$ is determined by the population $P_k$ in the current generation, the rate $r$ at which individuals reproduce and the level of available resources given that the system has a finite
capacity. According to May, such a one-dimensional map,
`even though simple and deterministic, can exhibit a surprising array of dynamical behaviour, from stable points, to a bifurcating hierarchy of stable cycles, to apparently random fluctuations' \cite{may1976simple}.

Consider the change of variables $x_k=rP_k-r/2$. This yields from (\ref{logistic}) the map
\begin{equation}\label{T-logistic}
    x_{k+1}=-x_k^2+a
\end{equation}
where $a =-r/2+r^2/4 \in \R$ is a scalar parameter.
The maps (\ref{logistic}) and
(\ref{T-logistic}) are topologically conjugate \cite{HK91}.
Comparing (\ref{T-logistic}) with (\ref{map FC}), it is seen that the
dynamics of an MC or an ML circuit in the FCD is exactly described by the logistic map (\ref{T-logistic}),
provided we choose $h\alpha I_0=a$ and $F(x)=(x+x^2)/(h\alpha)$. One possible solution to
these constraints is given by
\begin{equation}\label{choicesMCML}
    h=1,\ \ \alpha=1,\ \ F(x)=x^2+x,\ \ I_0=a.
\end{equation}

Summing up, under the conditions (\ref{choicesMCML}), the dynamics in the VCD of an MC or an ML circuit obeys the two-dimensional map (cf.\ (\ref{map-VCD}))
\begin{equation}\label{logistic-map-VCD}
\left\{
  \begin{array}{ll}
    y_{k+1}=y_k-(F(x_k+y_k)-F(x_k))=-y_k^2-2x_ky_k \\
    x_{k+1}=x_k+y_k.
  \end{array}
\right.
\end{equation}
The map (\ref{logistic-map-VCD}) has the first integral (cf.\ (\ref{integral MCML}))
$$
\Theta(x,y)= y+x+x^2
$$
and the two-dimensional state space $(x,y)$ is foliated in one-dimensional
invariant manifolds
\begin{align}
\label{Ma MCML both}
\cM(a)= \{ (x,y) \in \R^2: \Theta(x,y)=y+x+x^2=a \}
\end{align}
where $a \in \R$ is the manifold index.
On each manifold the dynamics is described in the FCD by the logistic map
\begin{equation}\label{map MCML FC}
    x_{k+1}=-x_k^2 +a.
\end{equation}

Next, we show that the two-dimensional map (\ref{logistic-map-VCD}) in the VCD displays an
extremely rich dynamics due to the foliation of its state
space and the property of extreme multistability. Moreover, we provide simulations
to illustrate and verify the predicted behavior.

\begin{figure}[ht]
  \centering%
  \includegraphics[width=.5\columnwidth]{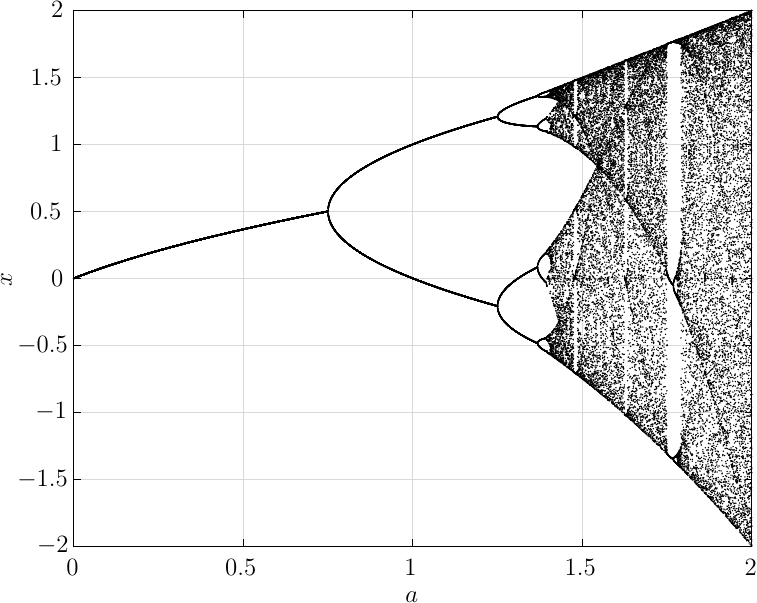}
\caption{Bifurcation diagram of the logistic map (\ref{T-logistic}) obtained by varying parameter $a$.}
\label{fig:BDlogi}
\end{figure}

\begin{figure}[ht]
  \centering%
  \includegraphics[width=0.5\columnwidth]{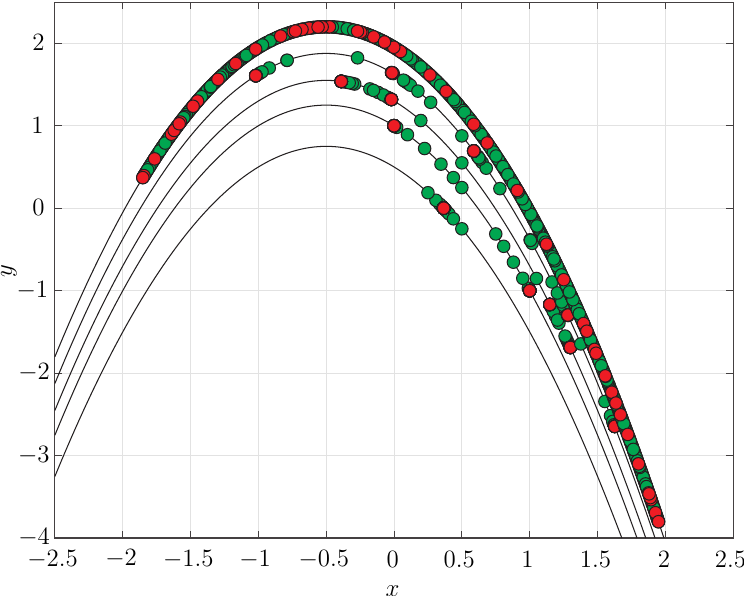}
\caption{Curves from bottom to top: (reverse) U-shaped invariant manifolds $\cM(a)$ of the
two-dimensional map (\ref{logistic-map-VCD}), see (\ref{Ma MCML both}), when $a \in \{0.5, 1, 1.3, 1.627, 1.95 \}$.
On each manifold the figure shows the evolution of
an orbit of (\ref{logistic-map-VCD}) starting on $\cM(a)$ (green circles), as well as the
long-term behavior of each orbit (red circles).}
\label{fig:manilogi}
\end{figure}

\begin{figure}[ht]
\begin{subfigure}{1 \columnwidth}
  \centering
  \includegraphics[width=0.5\linewidth]{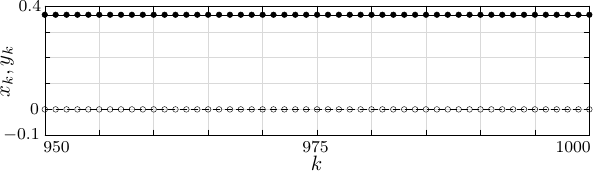}
  \caption{}
\end{subfigure}\\
\begin{subfigure}{1\columnwidth}
  \centering
  \includegraphics[width=0.5\linewidth]{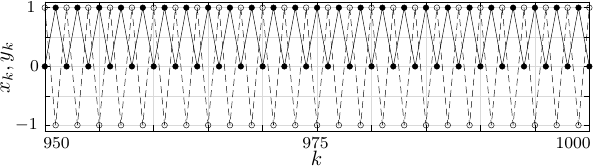}
  \caption{}
\end{subfigure}
\begin{subfigure}{1\columnwidth}
  \centering
  \includegraphics[width=0.5\linewidth]{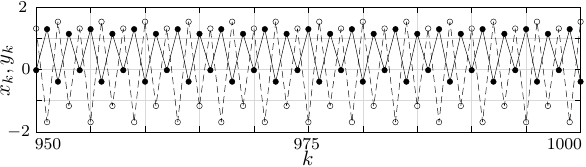}
  \caption{}
\end{subfigure}\\
\begin{subfigure}{1\columnwidth}
  \centering
  \includegraphics[width=0.5\linewidth]{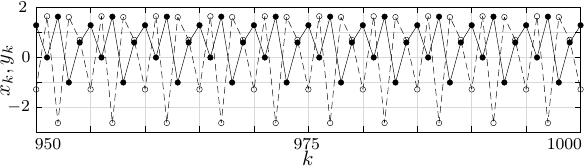}
  \caption{}
\end{subfigure}
\begin{subfigure}{1\columnwidth}
  \centering
  \includegraphics[width=0.5\linewidth]{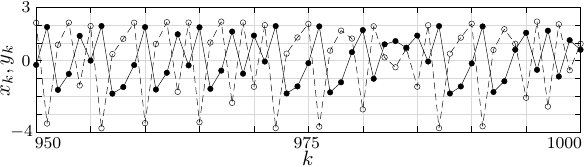}
  \caption{}
\end{subfigure}\\
\caption{Long-term behavior of the orbits of the two-dimensional map (\ref{logistic-map-VCD}) for
different values of the manifold index $a$. For better visualization,
the values $x_k$ (resp., $y_k$) are connected  by solid (resp., dashed) lines. (a) Fixed point for $a=0.5$, (b) period-two cycle
for $a=1$, (c) period-four cycle for $a=1.3$, (d) period-five cycle when $a=1.627$, (e) complex attractors when $a=1.95$.}
\label{fig:timedomainlogi}
\end{figure}

\begin{figure}[ht]
\centering
\begin{subfigure}{0.8\columnwidth}
  \centering
  \includegraphics[width=.5\linewidth]{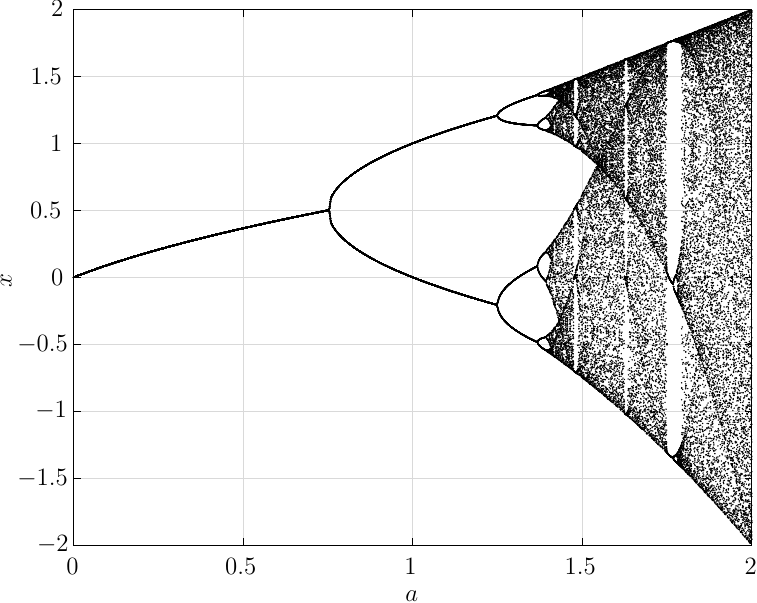}
  \caption{}
\end{subfigure}
\begin{subfigure}{0.8\columnwidth}
  \centering
  \includegraphics[width=.5\linewidth]{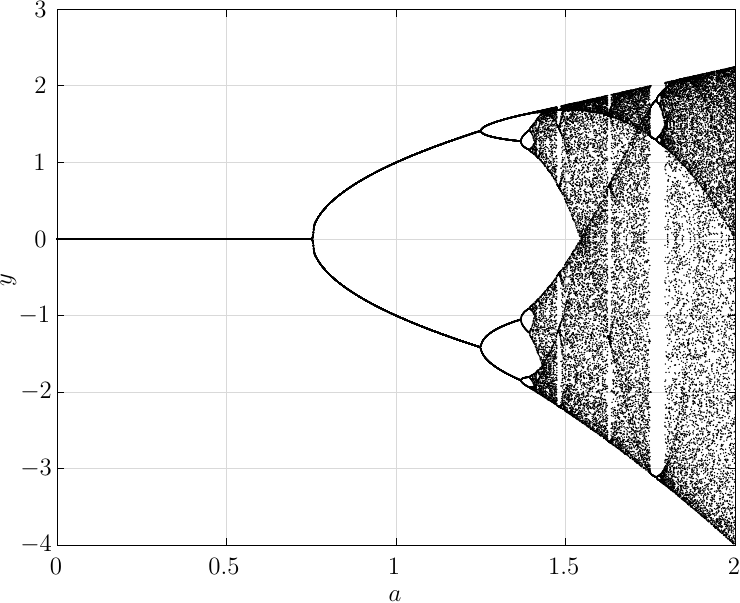}
  \caption{}
\end{subfigure}\\
\caption{a)
Bifurcation diagrams of the two-dimensional map (\ref{logistic-map-VCD}) obtained by varying parameter $a$: a) variable $x$ and b) variable $y$.}
\label{fig:BDlogixy}
\end{figure}

\begin{figure}[ht]
  \centering%
  \includegraphics[width=0.5\columnwidth]{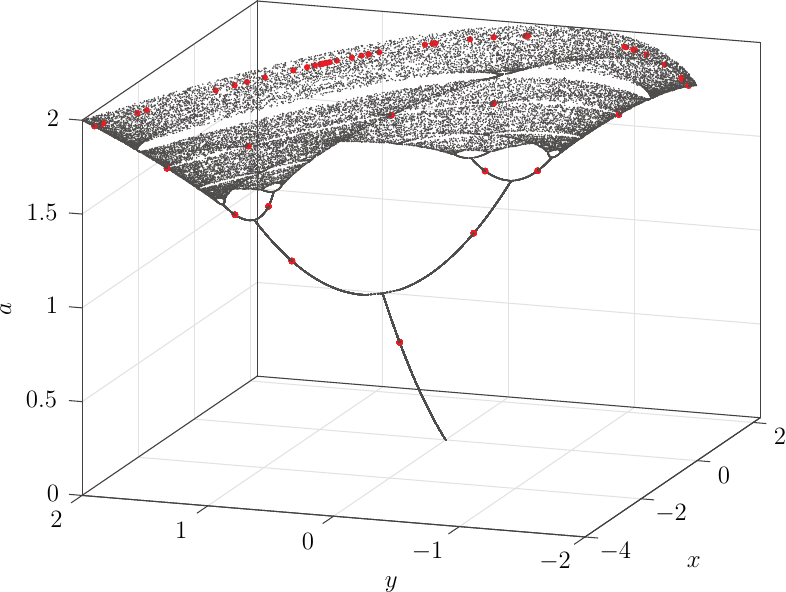}
\caption{A three-dimensional view of the bifurcation diagram of the two-dimensional map (\ref{logistic-map-VCD})
in the space $(x,y,a)$ obtained by varying parameter $a$.}
\label{fig:BDlogi3D}
\end{figure}

We have seen that on an invariant manifold with index $a$
the two-dimensional map (\ref{logistic-map-VCD}) embeds the logistic map
(\ref{map MCML FC}) with parameter $a$.
First, for convenience of exposition, let us briefly recall some basic
facts about the logistic map, see for instance \cite{hirsch2013differential}. The classic
bifurcation diagram of the logistic map is shown in Fig.\ \ref{fig:BDlogi}, where
we considered for simplicity the case $a \ge 0$.
When $0 \le a \le 2$ the orbits are bounded, while when $a>2$ orbits
escape to infinity. The orbits approach a stable fixed point when $0 \le a
\le a_1= 3/4$. At $a_1=3/4$, a flip (or period-doubling) bifurcation occurs with the birth
of a stable period-two cycle. By increasing $a$, we observe a second flip
bifurcation at $a_2=5/4$ where the period-two cycle gives birth to a
stable period-four cycle. This is followed by a typical cascade of period-doubling
bifurcations accumulating at $a_\infty =1.401 \dots$. These bifurcations originate
the birth of a complex attractor for $a>a_\infty$.
Let us now come back to the two-dimensional map (\ref{logistic-map-VCD}) in the VCD.
Its invariant manifolds (cf.\ \ref{Ma MCML both})
$$
y=a-x^2-x
$$
for some relevant values of $a$ are depicted in
Fig.\ \ref{fig:manilogi}. Note that they have a reverse U-shape due to the quadratic term $-x^2$. The same figure
shows for each manifold the transient part (green dots) and the attractor (red dots)
of an orbit starting on the manifold and evolving, due to its invariance, on the same
manifold.
When $a=0.5$ we have convergence to a fixed point, while when $a_1<a=1<a_2$ there is convergence to
a period-two cycle and when $a=1.3>a_2$ we observe convergence to
a period-four cycle. Moreover, when $a=1.627$ the orbit converges to a period-five cycle while when $a=1.95$ we observe convergence to a complex attractor.
The corresponding time-domain evolution of orbits for large $k$ is shown in
Fig.\ \ref{fig:timedomainlogi}.
Fig.\ \ref{fig:BDlogixy} displays the bifurcation diagram of $x$ and $y$ when varying
the manifold index $a$. Note that the bifurcation diagram of $x$ coincides with that of the logistic
map shown in Fig.\ \ref{fig:BDlogi}, confirming that the two-dimensional map (\ref{logistic-map-VCD}) embeds
as predicted the logistic map (\ref{T-logistic}). A three-dimensional view of the
bifurcation diagram of the map (\ref{logistic-map-VCD}) in the space $(x,y,a)$
is in Fig\ \ref{fig:BDlogi3D}.

\begin{remark}
It is important to stress that, \emph{given any parameter $a \in \R$,
the dynamics of the logistic map (\ref{T-logistic}) with parameter $a$
is embedded in the invariant manifold with index $a$
of the two-dimensional map (\ref{logistic-map-VCD}). Therefore,
all the different dynamics and attractors of the logistic map are embedded as a whole
in the set of invariant manifolds of the map (\ref{logistic-map-VCD}).}
This very rich behavior with infinitely many coexisting attractors
is an instance of the property of extreme multistability enjoyed by
maps of DT memristor circuits \cite{DiMarco20212577}. \qed
\end{remark}

\begin{remark}
Consider again the two-dimensional map (\ref{logistic-map-VCD}). It can be verified that the map
has a line of fixed points given by $(\bar y,\bar x)=(0,\chi)$, where $\chi$ is any value in $\R$.
The previous results can also be interpreted this way. Suppose that by changing the
index $a$ of the manifold we move along the line of fixed points. When $0\le a \le a_1$ the
positive fixed point
is stable. At $a=a_1$ the fixed point loses stability via
a flip bifurcation. We note that this
bifurcation happens for fixed parameters in the map (\ref{logistic-map-VCD}), hence we are
dealing with a so-called flip-bifurcation without parameters \cite{Corinto-Forti-II}. Similarly, we observe
a second flip bifurcation without parameters at $a=a_2$ and a subsequent cascade of flip
bifurcations without parameters for $a>a_2$. \qed
\end{remark}
\begin{remark}
It is possible to embed also other classic one-dimensional maps into
the two-dimensional map (\ref{logistic-map-VCD}). Consider for example the tent map \cite{hirsch2013differential}
\begin{equation}\label{tentmap}
    x_{k+1}=\left\{
              \begin{array}{ll}
                ax, & x<0.5 \\
                a(1-x), & x \ge 0.5
              \end{array}
            \right.
\end{equation}
where $a \in \R$ is a parameter.
This can be transformed into the logistic map (\ref{logistic}) via the homeomorphism
\cite{HK91}
$$
h(x)=\frac{2}{\pi} \arcsin(\sqrt{x})
$$
and hence the two maps are topologically conjugate.
Thereafter, we can proceed by the devised technique to embed the tent map into the
MC or the ML circuit (we omit the details). \qed
\end{remark}
\begin{remark}
There are intervals of parameters $a$ where chaos in the sense of Devaney and Li-Yorke
can be rigorously proved for the logistic map (\ref{T-logistic})
\cite{hirsch2013differential}. Since
the DT MC and ML circuits exactly embed on the invariant manifolds the logistic map,
it follows that we have constructed a class of memristor circuits where chaos can
be analytically proved. This differs from results in the literature where chaos
in DT memristor circuits is studied via numerical means as simulations, bifurcation
diagrams and Lyapunov exponents. \qed
\end{remark}

\section{Embedding the H\'enon and Lozi Map in the MLC and DMLC Circuit}
\label{sect:embed henon}
The celebrated Lorenz system is a three-dimensional system of ordinary differential equations modeling the convection and temperature variation of a hydrodynamical system.
The system, presented by Edward Lorenz in
a 1963 paper entitled `Deterministic Nonperiodic Flow' \cite{lorenz1963deterministic}, for suitable parameter values has a complex attractor and from simulations almost all its orbits appear to be of a nonperiodic nature. Later on, in the 1976 article entitled `A Two-dimensional Mapping with a Strange Attractor',
Michel H\'enon introduced a two-dimensional map constructed as a simplified model of the Poincare's section of the Lorenz system \cite{henon2004two} that inherits the stretch and folding mechanism leading to chaos in Lorenz system. The H\'enon map actually provides a more tractable mathematical model useful for better understanding Lorenz system. H\'enon map has been deeply studied and it is still the subject of intense investigation.

The two-dimensional H\'enon map is given as
\begin{equation}\label{Henon}
\left\{
  \begin{array}{ll}
    X_{k+1}= 1-a X_k^2 +Y_k \\
    Y_{k+1}=bX_k
  \end{array}
\right.
\end{equation}
where $a,b \in \R$ are two parameters. For suitable parameters
the map can display a chaotic attractor while for other parameter values it
shows a convergent or a periodic behavior.

\begin{figure}[ht]
  \centering%
  \includegraphics[width=.5\columnwidth]{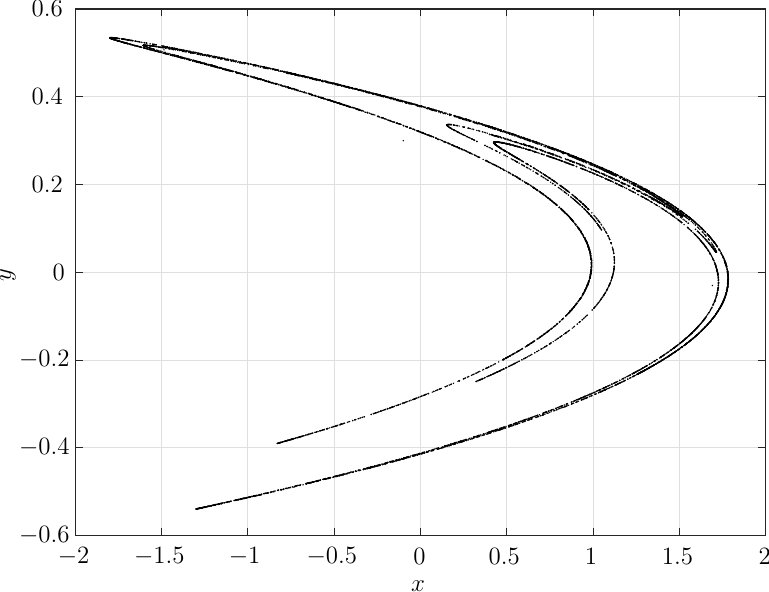}
\caption{Attractor displayed by H\'enon map (\ref{T-Henon}) when $a=1.4$ and $b=0.3.$}
\label{fig:Henon-attr}
\end{figure}

\begin{figure}[ht]
\centering
  \includegraphics[width=0.5\linewidth]{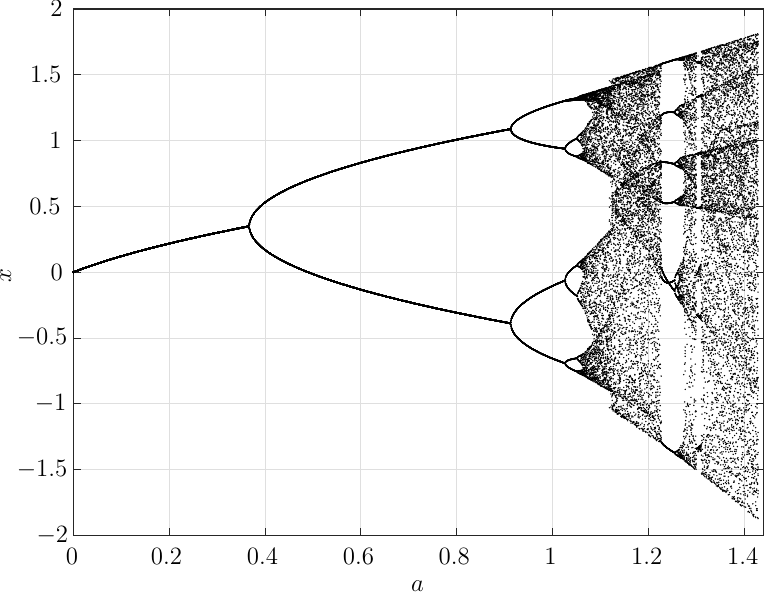}
\caption{Bifurcation diagram of variable $x$ for the H\'enon map (\ref{T-Henon}) obtained by varying parameter $a$ when $b=0.3$.}
\label{fig:BDHenonx}
\end{figure}

Let us consider the change of variables $x_k=aX_k$ and $y_k=aY_k$. We
obtain from (\ref{Henon}) the two-dimensional map
\begin{equation}\label{T-Henon}
\left\{
  \begin{array}{ll}
    x_{k+1}= -x_k^2 +y_k +a \\
    y_{k+1}=bx_k.
  \end{array}
\right.
\end{equation}
The two maps (\ref{Henon}) and (\ref{T-Henon})
are topologically conjugate.
As an example, Fig.\ \ref{fig:Henon-attr}
shows the chaotic attractor with a boomerang-like shape of H\'enon map (\ref{T-Henon})
obtained for the traditional parameters $a=1.4$ and $b=0.3$ considered in \cite{henon2004two}, while
Fig.\ \ref{fig:BDHenonx} shows the bifurcation diagram of $x$ when parameter
$a$ is varied and $b=0.3$.

Comparing (\ref{bothMLC-FCD}) with (\ref{T-Henon}), it is easy to see that the dynamics
of an MLC or a DMLC circuit is exactly described in the FCD by the H\'enon map (\ref{T-Henon}),
provided we choose $h\alpha=1$, $h\beta=1$, $h\gamma=b$, $F(x)=x+x^2$ and $I_0=a$. One possible
solution is given by
\begin{equation}\label{choicesMLC}
    h=1,\ \ \alpha=\beta=1,\ \ \gamma=b,\ \ F(x)=x^2+x,\ \ I_0=a.
\end{equation}

We can wrap up these results as follows.
Under the conditions (\ref{choicesMLC}), the dynamics in the VCD of an MLC or a DMLC
circuit obeys the three-dimensional map (cf.\ (\ref{bothMLC-VCD}))
\begin{equation}\label{Henon-map-VCD}
\left\{
  \begin{array}{ll}
w_{k+1}=w_k-(F(x_k+w_k)-F(x_k))+z_k \\
       =-w_k^2-2x_kw_k+z_k\\
    z_{k+1}=bw_k \\
    x_{k+1}=x_k+w_k.
  \end{array}
\right.
\end{equation}
The map (\ref{Henon-map-VCD}) has the first integral (cf.\ (\ref{integral MLC}))
$$
\Theta(x,w,z)= w+z-bx+x+x^2
$$
and the three-dimensional state space $(x,w,z)$ is foliated in one-dimensional
invariant manifolds
\begin{align}
\label{Mani Henon}
\cM(a)= \{ (x,w,z) \in \R^3: \Theta(x,w,z)&=w+z\\
-bx+x+x^2=a \}
\end{align}
where $a \in \R$ is the manifold index.
It turns out that on each manifold the dynamics is described in the FCD by the H\'enon map
(cf.\ \ref{bothMLC-FCD})
\begin{equation}\label{MAP FCD MLC}
\left\{
  \begin{array}{ll}
    x_{k+1}= -x_k^2 +y_k +a \\
    y_{k+1}=bx_k.
  \end{array}
\right.
\end{equation}

\begin{figure}[ht]
\centering
  \includegraphics[width=0.5\linewidth]{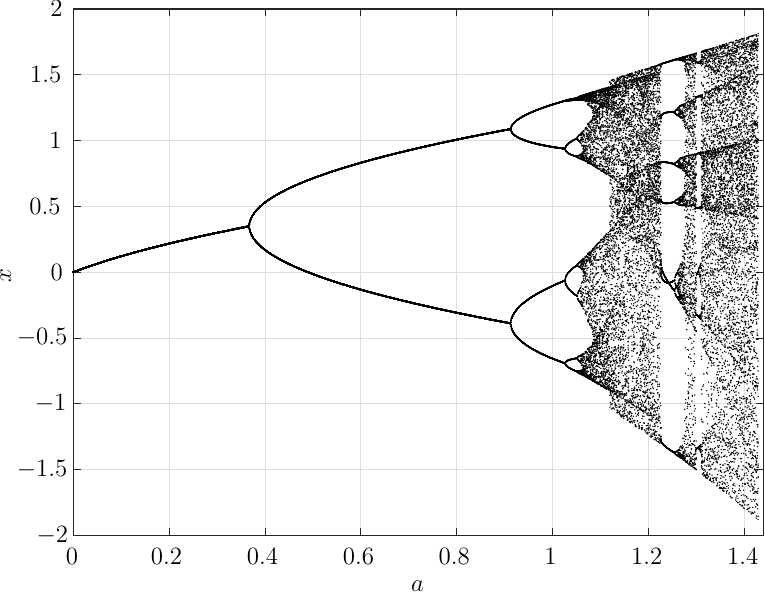}
\caption{Bifurcation diagram of variable $x$
three-dimensional map (\ref{Henon-map-VCD}) obtained
by varying parameter $a$ when $b=0.3$.}
\label{fig:BD henon 3d x}
\end{figure}

The three-dimensional map (\ref{Henon-map-VCD}) in the VCD displays an
extremely rich dynamics due to the foliation of its state
space and the property of extreme multistability. Fig.\ \ref{fig:BD henon 3d x}
shows the bifurcation diagram of variables $x$ of (\ref{Henon-map-VCD})
obtained by varying $a$ when $b=0.3$. This coincides with that in
Fig.\ \ref{fig:BDHenonx}, confirming that (\ref{Henon-map-VCD}) embeds
the H\'enon map for the considered parameters and memristor nonlinearity.

\begin{remark}
Once more we stress that, \emph{given any parameter $a \in \R$,
the dynamics of the H\'enon map (\ref{T-Henon}) with parameter $a$
is embedded in the invariant manifold with index $a$
of the three-dimensional map (\ref{Henon-map-VCD}). Therefore,
all the different dynamics and attractors of the H\'enon map are embedded as a whole
in the set of invariant manifolds of the map (\ref{Henon-map-VCD}).}
This extremely rich behavior with infinitely many coexisting attractors
is once more a consequence of the property of extreme multistability of DT maps
implemented by memristor circuits \cite{DiMarco20212577}. \qed
\end{remark}

\begin{remark}
\label{rem:bpositive}
If we eliminate the controlled sources in the MLC or DMLC, i.e., we put
$\mu=0$ or $\nu=0$, respectively, it is seen that the simplified
MLC and DMLC circuits can implement a variant
of H\'enon map where parameter $b<0$. Such a variant is still able to display
interesting complex behaviors as discussed for instance in \cite{PhysDhenon}. \qed
\end{remark}

\begin{remark}
In 1978, in a short note entitled `Un attracteur \'etrange (?) du type attracteur de H\'enon' \cite{lozi1978attracteur},
Ren\'e Lozi proposed [1] the following two-dimensional map
\begin{equation}\label{Lozin}
\left\{
  \begin{array}{ll}
    x_{k+1}= 1-a |x_k| +y_k \\
    y_{k+1}=bx_k
  \end{array}
\right.
\end{equation}
where $a,b \in \R$ are two parameters. The defining equations
are exactly those of the H\'enon map, provided
a quadratic term in H\'enon map is replaced with a piecewise linear term in Lozi
map. This permits
to rigorously prove the chaotic character of some attractors
and to arrive at a detailed analysis
of attraction basins. In his original paper, Lozi used the parameters $a=1.7$
and $b=0.5$ to obtain a complex attractor with an edgy boomerang-like shape
whose chaotic nature was rigorously
proved by Michal Misiurewicz \cite{misiurewicz1980strange}. A review on Lozi maps and its applications is
provided in \cite{elhadj2013lozi,lozi2023survey}. Again, it is not difficult to show that
we can embed Lozi map in an MLC circuit or its dual by appropriately choosing the
parameters and the piecewise linear memristor nonlinearity (details are left
to the interested reader).
\qed
\end{remark}

\section{Conclusion}
\label{sect:concl}
One relevant fact that emerges in the literature is that maps obtained
by discretizing simple memristor circuits are able to display via simulations an unexpectedly rich dynamic behavior including chaos and hyperchaos. This paper has provided a rationale to explain this complex dynamics via the analysis of two simple classes of DT memristor circuits obtained via DT-FCAM.
Firstly, a DT circuit composed of a capacitor plus a flux-controlled memristor and its dual, have been considered. It is shown that the two-dimensional map in the VCD of the circuits can exactly embed
on the invariant manifolds, i.e., in the FCD, two classic chaotic one-dimensional maps, i.e., the logistic and the tent map. Secondly, a DT memristor Murali-Lakshmanan-Chua and its dual, have been considered. The paper has shown that the three-dimensional map in the VCD of the circuits can exactly embed
on the invariant manifolds two classic chaotic two-dimensional maps, i.e., the H\'enon and the
Lozi map. One significant result is that we can exploit the property of extreme multistability
of the DT memristor circuits to simultaneously embed in the set of invariant manifolds
all possible dynamics obtained by varying one parameter in the logistic or in the
H\'enon map. The main conclusion is that observing complex dynamics
even in maps obtained by discretizing  the simplest memristor circuits is not surprising
at all.

\end{document}